\documentclass[letter,11pt]{article}
\usepackage{graphicx}
\usepackage[colorlinks=true,linkcolor=black,urlcolor=black,citecolor=black]{hyperref}
\usepackage{caption}
\usepackage{amssymb}
\usepackage{amsmath}
\usepackage{bbm}
\usepackage{outlines}
\usepackage{geometry}
\usepackage{enumitem}
\usepackage{parskip}
\usepackage{float}
\usepackage{natbib}
\usepackage{setspace}
\usepackage{authblk}
\usepackage[noabbrev,capitalise]{cleveref}
\creflabelformat{equation}{#2\textup{#1}#3} 

\newcommand{\Rho}{\mathrm{P}}
\newcommand{\E}{\mathbb{E}}

\newcommand{\T}{^{\textup{T}}}
\newcommand{\R}{\mathbb{R}}
\newcommand{\norm}[1]{\left\lVert#1\right\rVert}

\title{Reduced-rank generalized bilinear models}

\author[1]{Kevin S. Kapner\footnote{Corresponding author, E-mail: kkapner@fas.harvard.edu}}
\author[1]{Jeffrey W. Miller}
\affil[1]{Department of Biostatistics, Harvard T.H. Chan School of Public Health}

\date{}

\begin{document}

\doublespacing

\maketitle

\begin{abstract}
Dimensionality reduction and effect estimation are central tasks in the analysis of high-dimensional data such as in genomics.
Generalized bilinear models (GBMs) provide a versatile framework for these tasks, however, the statistical and computational efficiency of GBMs degrades rapidly as the number of sample covariates grows. To address this limitation, we introduce reduced-rank generalized bilinear models (RR-GBMs), which employ a reduced-rank sample coefficient matrix to model the effect of a large number of covariates without requiring an excessive number of parameters.  In simulation studies, we find that when the true sample coefficient matrix is reduced rank or close to reduced rank, RR-GBM outperforms the standard full-rank GBM both statistically and computationally, providing more accurate estimates with lower computational burden. We develop a data thinning approach for model selection in the RR-GBM framework, facilitating rank selection.  Furthermore, RR-GBM enables a new approach to visualizing the relationships among covariates and among features. We demonstrate the method in an application to Perturb-seq data for pancreatic cancer.
\end{abstract}

\section{Introduction}

In single-cell RNA sequencing (scRNA-seq) experiments, researchers often have a large number of sample covariates describing the cells.
Some covariates, such as batch, represent effects that one would like to adjust for, since they are not of direct interest.
Other covariates, such as experimental condition, represent effects of primary interest that one would like to infer.
Modern protocols often produce data across many experimental conditions, for example, Perturb-seq performs scRNA-seq on cells with distinct targeted mutations induced by a gene editing technology like CRISPR \citep{dixit_perturb-seq_2016, adamson_multiplexed_2016, southard_comprehensive_2025}. In these experiments, researchers aim to identify the effect of each perturbation, such as gene loss, expression amplification, or expression reduction, on a phenotype of interest such as survival. 

For Gaussian-distributed matrix data, factor analysis and other bilinear models are commonly used to perform dimensionality reduction and effect estimation \citep{takane_principal_1991, tipping_probabilistic_1999}.
Meanwhile, for non-Gaussian matrix data, generalized bilinear models (GBMs) extend bilinear models by employing a generalized linear model (GLM) for the entries \citep{choulakian_generalized_1996, townes_feature_2019, gabriel_generalised_1998, miller_inference_2020}.

This flexible approach models the effect of sample (e.g., cell) covariates, feature (e.g., gene) covariates, and latent factors in an integrated statistical framework.
However, the inclusion of sample covariates in a GBM presents a challenge in high-dimensional data, since each additional sample covariate requires estimating a parameter for every feature \citep{miller_inference_2020}.
Consequently, for genomics data where the number of features can exceed $20{,}000$, very few sample covariates can be included, otherwise the estimation accuracy and computational burden become unacceptable.
In the multivariate regression setting---without feature covariates or latent factors---this problem is dealt with using reduced-rank regression \citep{izenman_reduced-rank_1975, yee_reduced-rank_2003, camba-mendez_tests_2003, davies_procedures_1982, braak_biplots_1994}, but these ideas have not yet been explored in the GBM setting.

In this paper, we introduce the reduced-rank generalized bilinear model (RR-GBM) and provide a computationally tractable estimation algorithm that scales to large data sets. Our method estimates a rank-reduced version of the coefficient matrix for the sample covariates, resulting in a significant reduction in the number of parameters in the high-dimensional setting. RR-GBM outperforms the standard full-rank GBM when the true coefficient matrix is approximately reduced-rank or when uninformative covariates are included in the model. The method also provides a way to quantify variable importance and visualize structure among the sample covariate effects via low-dimensional summaries. To select the rank of the sample covariate coefficient matrix as well as the rank of the latent factor matrix, we provide a data-driven model selection method based on data thinning \citep{neufeld_inference_2024}.

The article is organized as follows.
In \cref{sec:methodology}, we introduce the RR-GBM methodology.
\cref{sec:simulations} contains simulation studies illustrating various aspects of the method.
\cref{sec:application} presents an application to Perturb-seq type data.
\cref{sec:discussion} concludes with a brief discussion.

\section{Methodology}
\label{sec:methodology}

\subsection{Reduced-rank generalized bilinear model}

Suppose we have a random data matrix $Y \in \R^{I\times J}$, a matrix of feature covariates $X \in \R^{I\times K}$, and a matrix of sample covariates $Z \in \R^{J\times L}$.  Generalized bilinear models (GBMs) assume the form
\begin{align}
\label{eq:gbm}
g(\E Y) = X A\T + B Z\T + X C Z\T + U \Sigma V\T
\end{align}
where $A \in \R^{J \times K}$, $B \in \R^{I\times L}$, and $C \in \R^{K \times L}$ are matrices of coefficients, $U \Sigma V\T$ is a low-rank matrix of latent effects, and $g(\cdot)$ is a known link function that is applied element-wise; it is assumed that $g(\cdot)$ is a smooth function such that the derivative $g'$ is positive.
Here, $Y$, $X$, and $Z$ are observed, whereas $A$, $B$, $C$, $U$, $\Sigma$, and $V$ are unknown parameters to be estimated.  
The interpretation of the dimensions is that $I$ is the number of features, $J$ is the number of samples, $K$ is the number of feature covariates, $L$ is the number of sample covariates, and $M$ is the number of latent factors, which may also need to be estimated. 
All of the parameters are identifiable under certain constraints.
See \citet{miller_inference_2020} for an in-depth exposition of these models.

We propose to constrain $B$ to be a lower-rank matrix, $B = Q \Lambda R\T$ yielding a \emph{reduced-rank generalized bilinear model} (RR-GBM) in which
\begin{align}
\label{eq:rrgbm}
g(\E Y) = X A\T + Q \Lambda R\T Z\T + X C Z\T + U \Sigma V\T
\end{align}
where $Q \in \R^{I \times N}$ and $R \in \R^{L \times N}$ are orthonormal frames, $\Lambda \in \R^{N \times N}$ is diagonal, and $N$ represents the desired rank of $B$.
The rest of the matrices are defined the same way as in the standard GBM.
All of the parameters of the RR-GBM are identifiable---including $Q$, $R$, and $\Lambda$---under natural corresponding constraints, as we discuss in \cref{sec:interpretation}.

The distribution of $Y_{ij}$ is taken to be an exponential dispersion family. More specifically, in this paper, it is taken to be either the Poisson or the Negative Binomial family.

\subsection{Interpretation of the model}
\label{sec:interpretation}

The motivation for the RR-GBM is that in the standard GBM, the total number of parameters grows as approximately $I L$ as the number of sample covariates $L$ increases.  Since $I$ is typically large, on the order of $10^3$--$10^4$, the computational and statistical efficiency of estimating a full rank $B$ matrix deteriorates quickly as $L$ increases.
In contrast, by using a reduced rank matrix $B = Q \Lambda R\T$ in the RR-GBM, the total number of parameters grows as approximately $N L$ as a function of $L$.  Since $N$ is typically small, for instance, $N \leq 10$ or $20$, this yields a massive decrease in the number of parameters that must be estimated.
Indeed, in the common use settings described above, the total number of parameters is reduced  by a factor of approximately $N/L$; see \cref{sec:number-of-parameters} for the derivation and exact reduction factor. 

The components of the decomposition $B = Q\Lambda R\T$ each reflect a different facet of the structural relationship between the sample covariates and the outcomes. First, the $n$th column of $R$ represents the effect of each covariate on the latent value of factor $n$.  
The $n$th diagonal entry of $\Lambda$ represents the overall magnitude of the effect that factor $n$ has on the outcome.
The $n$th column of $Q$ represents the effect that factor $n$ has on each feature (e.g., gene).   Putting these pieces together, we obtain a mechanistic representation of how each sample covariate affects the outcome for a given feature in a given sample.
The intuition is that the effect of the covariates is mediated by a low-dimensional space of $N$ latent values.

The parameters are not identifiable unless we place appropriate constraints on them. To this end, we assume all of the constraints specified by \citet{miller_inference_2020}, detailed in \cref{sec:complete-ident-constraints} for completeness, as well as additional constraints pertaining to $Q$, $\Lambda$, and $R$. Specifically, the additional constraints we assume for identifiability of the RR-GBM are:
\begin{enumerate}[label=(\alph*)]
    \item $\Lambda$ is a diagonal matrix such that $\lambda_{11} > \lambda_{22} > \cdots > \lambda_{NN} > 0$,
    \item $Q\T Q = \mathrm{I}$ and $R\T R = \mathrm{I}$,
    \item $X\T Q\Lambda R\T = 0$, and
    \item the first nonzero entry of each column of $Q$ is positive,
\end{enumerate}
where $Q \in \mathbb{R}^{I \times N}$, $\Lambda \in \mathbb{R}^{N \times N}$, $R \in\mathbb{R}^{L \times N}$, and $N < \min\{I, L\}$.

Appropriate selection of the rank of a factorization is a perennial problem, since there is often no clear choice based on fit to the training data, and it is not obvious  how to hold out a test set when the data set is a matrix.
We present a method for selecting $N$ and $M$, the ranks of $Q\Lambda R\T$ and $U\Sigma V\T$, respectively, using the data thinning technique of \citet{neufeld_inference_2024} to split the data matrix into independent training and test matrices; see \cref{sec:count-splitting}.

\subsection{Estimation algorithm for RR-GBM}
\label{sec:estimation}

One approach to fitting the RR-GBM would be to first estimate a full-rank $B$ matrix using an existing GBM algorithm, and then apply the singular value decomposition (SVD) to factorize $B = Q\Lambda R\T$. However, this is not equivalent to enforcing the reduced-rank constraint during estimation by jointly estimating $Q$, $\Lambda$, and $R$ along with the rest of the parameters, 
which we find yields better performance. In this section, we describe our joint estimation algorithm.

In our proposed algorithm for RR-GBM, the update steps for $A$, $C$, $\Sigma$, $U \Sigma$, and $\Sigma V\T$ are the same as those of \citet{miller_inference_2020}. Our algorithm differs only in that we estimate $Q$, $\Lambda$, and $R$ instead of estimating $B$. The updates involving $Q$, $\Lambda$, and $R$ are done in three sequential steps: (1) updating $Q\Lambda$, (2) updating $\Lambda R\T$, and (3) updating $\Lambda$.

Briefly, updating $Q\Lambda$ consists of bounded regularized Fisher scoring steps for the product $Q\Lambda$, followed by the enforcement of identifiability constraints, with the updated $Q\Lambda R\T$ coming from a truncated SVD of the updated $B$ matrix.
Updating $\Lambda R\T$ is done in the same way, using steps derived for $\Lambda R\T$ instead of $Q\Lambda$.
One difference, however, is that updating $\Lambda R\T$ requires the computation of a block matrix of dimension $IL \times IL$ with block sizes of $I \times I$, which can be computationally prohibitive when $I$ and $L$ are large.
Thus, to speed up computation of the block Fisher scoring steps for $\Lambda R\T$, we use multi-threaded linear algebra routines.  Updating $\Lambda$ consists of a bounded regularized Fisher scoring step. We iteratively cycle through all of the update steps above until a convergence criteria is met or a maximum number of iterations has occurred. See \cref{sec:estimation-algo} for a step-by-step description of the full estimation procedure.

\subsection{Visualizing the effect of the sample covariates}
\label{sec:vis-sample-cov}
An important feature of factorization methods is that they enable one to visualize high-dimensional data in an interpretable way via dimensionality reduction. For instance, principal components analysis (PCA) is often used to visualize clusters and other structures in a collection of high-dimensional data points.
Similarly, in RR-GBM, the factorization $B = Q\Lambda R\T$ can roughly be thought of as an application of PCA to the sample covariate coefficient matrix $B$, and thus, it is amenable to the same visual interpretations.

To illustrate this point, we can visualize the effect of each covariate on each latent factor value by making a heatmap of the estimated $\Lambda R\T$ matrix; see \cref{fig:sample-covariate-visualization,fig:covar-struct,fig:phenotypic-screen} for examples.
Furthermore, for any two latent factor dimensions, we can make a scatterplot of these effects by plotting the corresponding rows of the estimated $\Lambda R\T$ matrix; see \cref{fig:phenotypic-screen}. This is analogous to familiar scatterplots of the first two PC scores, except that each point in the plot corresponds to a covariate rather than a sample; hence, this visualization technique may reveal clusters of covariates or other relationships among covariates in terms of their effect on the outcomes.

Likewise, we can visualize the effect of each latent factor on each feature (e.g., gene) by making a heatmap of the $Q$ matrix, or alternatively, making a scatterplot based on any two columns of $Q$; see Figure \ref{fig:phenotypic-screen}.  In such a scatterplot, each point in the plot corresponds to a feature; hence, this may reveal clusters or other relationships among features in terms of how they are affected by the corresponding factors.

\subsection{Rank selection using data thinning}
\label{sec:count-splitting}

In this section, we describe our proposed technique for selecting $N$, the rank of $B = Q \Lambda R\T$. The same approach can also be used to select the rank of $U \Sigma V\T$.
Ideally, one would choose $N$ to minimize the distance between the estimate $\hat{B}$ and the true matrix $B$, but of course the true $B$ is unknown in practice. 
For situations with i.i.d.\ samples, one can split the data into training and test sets, and quantify model performance by fitting on the training set and measuring prediction performance on the test set.

However, it is not appropriate to assume the samples are i.i.d.\ in our setting.
In particular, since the model involves sample-specific parameters, it is not possible to fit on one set of samples and test on another.

To remedy this problem, we propose using data thinning to split the original data matrix into a training data matrix and test data matrix, both having the same dimensions as the original matrix \citep{neufeld_inference_2024}. Here, we focus on the case of Poisson outcomes, but the technique generalizes to many other common distributions as well.
Suppose we have a random data matrix $Y \in \mathbb{R}^{I\times J}$ where $Y_{ij} \sim \mathrm{Poisson}(\mu_{ij})$. For all $i$ and $j$, given $Y_{i j}$, sample $Y'_{ij} \sim \mathrm{Binomial}(Y_{ij}, 1/2)$ and set $Y''_{ij} = Y_{ij} - Y'_{ij}$. It can be shown that $Y'_{ij}$ and $Y''_{ij}$ are independent $\mathrm{Poisson}(\mu_{i j}/2)$ random variables, when $Y_{i j}$ is marginalized out \citep{neufeld_inference_2024}. 
Thus, the matrices $Y'$ and $Y''$ are independent data matrices with half the mean of the original data matrix. For each candidate rank $N$, we use $Y'$ to fit the model parameters and then generate our predictions $\widehat{\E} Y$ using the covariate values.  The RMSE between $\widehat{\E} Y$ and $Y''$ is then obtained. Plotting these RMSE values versus the ranks of the fitted model yields a scree-like plot that helps guide rank selection; see \cref{fig:empirical-rank-selection}.

\section{Simulations}
\label{sec:simulations}

In this section, we evaluate the RR-GBM methodology across a range of simulation studies.
In each simulation, we randomly generate true parameters for an RR-GBM and simulate a random dataset $Y$ using these true parameters, with either a Poisson or Negative Binomial outcome distribution for $Y_{ij}$ depending on the simulation; see \cref{sec:sim-data-generation} for full details. 

\subsection{Convergence of RR-GBM estimates to true parameter values}

To assess the convergence of our parameter estimates to the true values, we performed a simulation study with a range of sample sizes $J$ and sample coefficient matrix ranks $N_{0}$. Specifically, for each $J \in \{500, 1000, 2000, 5000\}$ and $N_{0}\in\{1,2,3,4,5, 6\}$, we simulate Poisson data $Y_{i j}$ (see \cref{sec:sim-data-generation}), fit the RR-GBM, and compare the estimated parameters to the true parameters. Here, we use $I = 20{,}000$ features, one latent factor ($M = 1$), and 6 sample covariates ($L = 6$) in both the true and estimated RR-GBMs.

\begin{figure}
    \centering
    \includegraphics[width=0.8\linewidth]{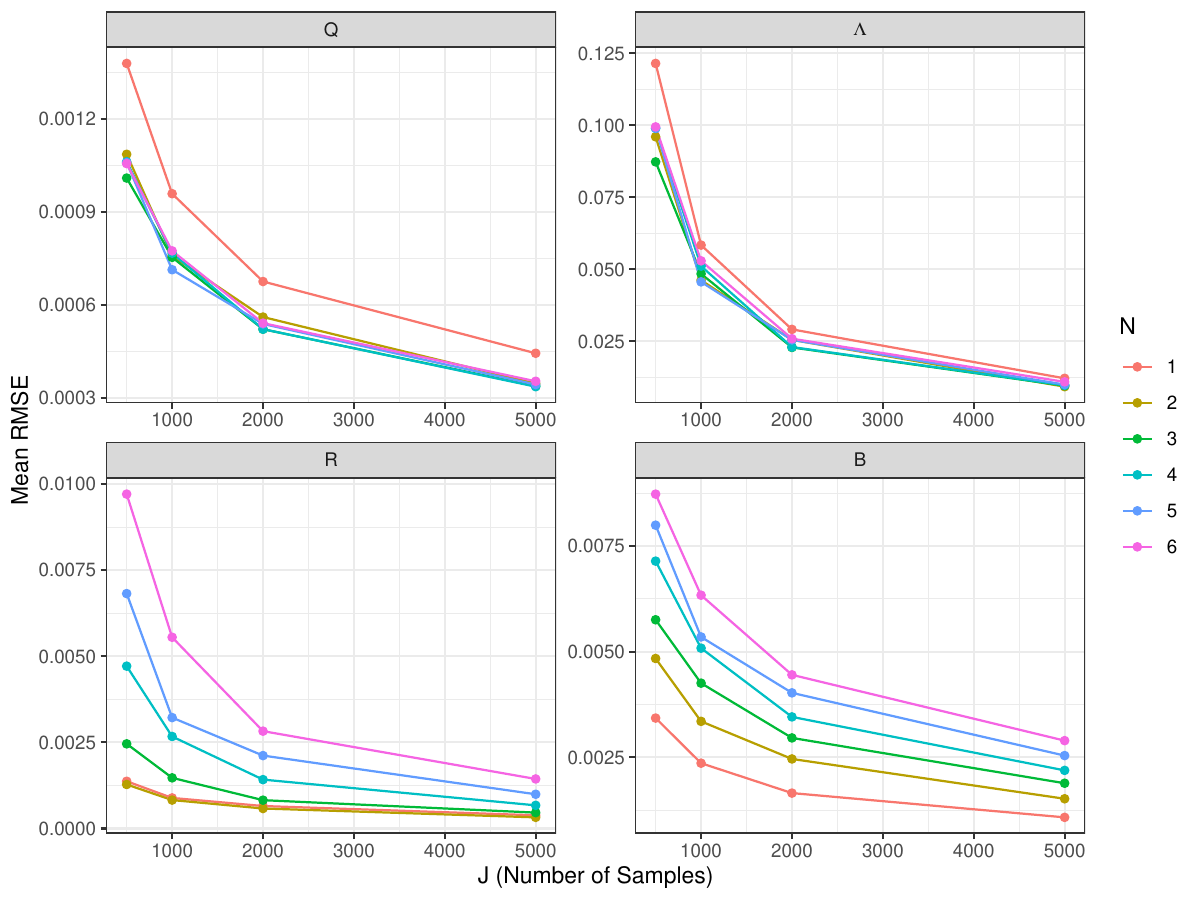}
    \caption{\textbf{Convergence of RR-GBM estimates.} RMSE between the true and estimated matrices for $Q$, $\Lambda$, $R$, and $B$ versus increasing sample size, in the case of Poisson outcomes. In the bottom right plot, $B = Q\Lambda R\T$ and the estimate is $\hat{B} = \hat{Q}\hat{\Lambda}\hat{R}\T$. The true rank $N$ of $B$ was used for fitting the RR-GBM in each case.} 
    \label{fig:sample-feature-simulation}
\end{figure}

For each dataset, we fit the model using the estimation algorithm described in \cref{sec:estimation}, setting the model rank $N$ to the true rank used to generate the data.
We evaluate performance by computing the mean RMSE between the true $B$ and estimate $\hat{B}$. In \cref{fig:sample-feature-simulation}, we observe that RMSE decreases as the sample size grows, as expected. We also observe that higher ranks $N$ lead to higher RMSE between $\hat{B}$ and $B$, due to the larger number of parameters that must be estimated. 

\subsection{Performance of RR-GBM estimates in the reduced-rank setting}

Next, we evaluate the effect of the rank $N$ and number of covariates $L$ on estimation accuracy.
We consider two scenarios: (i) the true $B$ matrix is exactly reduced rank, and (ii) the true $B$ matrix is approximately reduced rank. The exactly reduced-rank setting is when the true $B$ matrix has rank $N_{0} < L$, which occurs when the effect of the covariates is mediated by $N_{0}$ variables that are linear combinations of the covariates.
A special case of this would be when only $N_{0}$ covariates have nonzero effects.
The approximately reduced-rank setting is when the true $B$ matrix can be closely approximated by a rank $N_{0} < L$ matrix, which occurs when the effects are predominantly captured by $N_{0}$ factors.

In the full GBM, a total of $(I-K) L$ parameters need to be estimated for $B$, but in the RR-GBM, only $N (I-K+L-N)$ parameters need to be fit for $B$. For given $I$, $K$, and $L$ with $I \gg \max\{K,N\}$, which is the case for scRNA-seq data, a full GBM requires fitting around $L/N$ times as many parameters for $B$ as a rank $N$ RR-GBM; see \cref{subsubsec:reduc-in-param}.

Thus, we expect that as the number of sample covariates $L$ grows, the benefit of using an RR-GBM will increase.

\subsubsection{Exact reduced-rank setting}

When the true rank of $B$ is $N_{0}$, we observe that the estimation error for $B$ decreases as we increase $N$ (the rank of $B$ in the fitted RR-GBM), until $N = N_{0}$ (\cref{fig:true-reduced-rank-performance}). If a model of rank $N > N_{0}$ is fit, the estimation procedure terminates without finding a solution. This occurs because when the true $B$ has rank less than $N$, there is not a way to obtain a unique rank $N$ approximation of $B$, so the compact SVD does not have a solution. This behavior of the procedure can be used to quickly identify any covariates that have linearly dependent coefficients, that is, covariates that have the same effect on all of the outcomes.

\subsubsection{Approximate reduced-rank setting}

\begin{figure}
    \centering
    \includegraphics[width=1\linewidth]{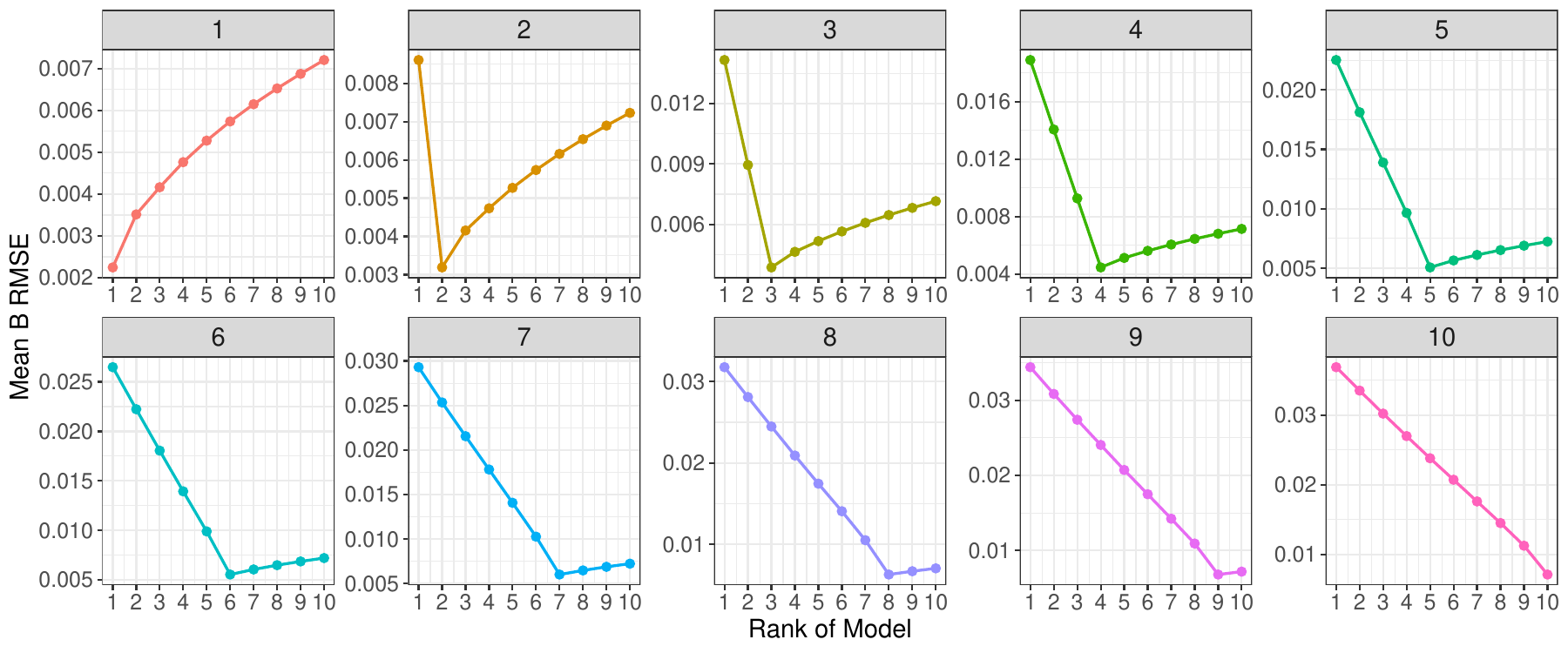}
    \caption{\textbf{RR-GBM estimation accuracy as a function of true rank and model rank.}
    Panels are labeled with their respective true approximate rank of $B$ (corresponding $L-N_{0}$ singular values are 0.01 times their randomly generated value). The average RMSE (over 100 simulations) between the estimated $B$ (given by $\hat{Q}\hat{\Lambda}\hat{R}\T$) and the true $B$ are displayed. Data was generated using a Negative Binomial outcome distribution and fit with a model with 1 latent factor ($M =1$).}
    \label{fig:reduced-rank-performance}
\end{figure}

When the true $B$ matrix is approximately rank $N_{0}$, we observe in \cref{fig:reduced-rank-performance} that the estimation error for $B$ decreases as we increase $N$ until $N = N_{0}$, and then the error increases again until $N=L$.  The maximum possible value of $N$ is $L$ since there are $L$ covariates; the RR-GBM with $N = L$ is equivalent to the standard full GBM \citep{miller_inference_2020}.

\begin{figure}
    \centering
    \includegraphics[width=0.8\linewidth]{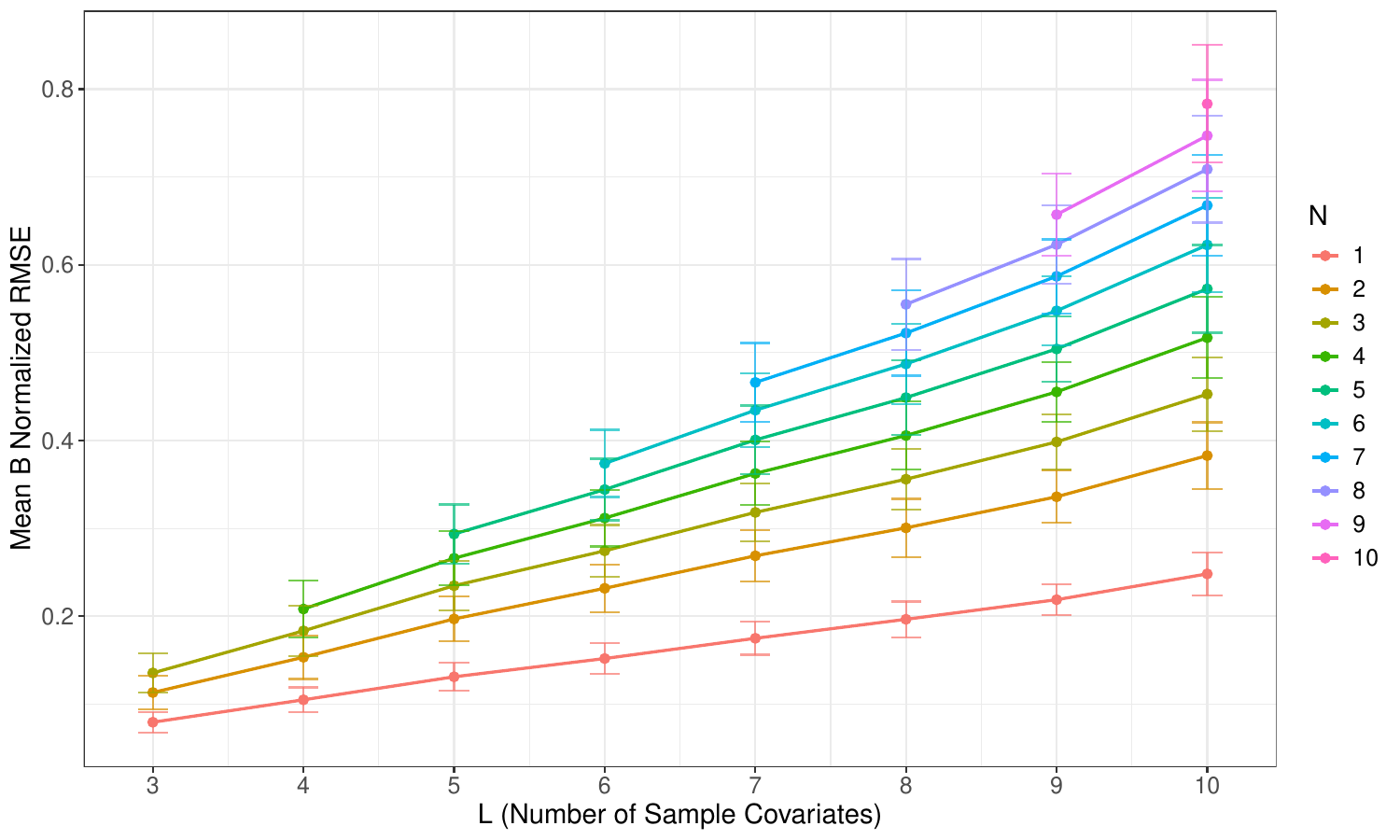}
    \caption{\textbf{Performance of RR-GBM estimates as $L$ increases.} Normalized RMSE between the true and estimated $B$ plotted against the number of sample covariates $L$, for Negative Binomial outcomes with one latent factor ($M=1)$, $J = 2500$, $I = 5000$, and $K=1$. The approximate rank of $B_{0}$ was $N_0 = 1$ for all $L$. Each point represents the average over 100 simulation runs and the error bars denote standard deviation. Normalized RMSE is defined as $\lVert \hat{B} - B_{0}\rVert_{F} / \lVert B_{0} \rVert_{F}$ where $\lVert \cdot \rVert_{F}$ denotes the Frobenius norm.  This normalization was performed to adjust for the increasing size of $B_{0} \in \mathbb{R}^{I\times L}$ so that performance can be compared across $L$. We see that $N = 1$ always has the smallest normalized RMSE across increasing $L$ and that normalized RMSE increases monotonically with an increasing fitted $N$. The top point for each $L$ (corresponding to $N = L$) is equivalent to the full GBM model and always has the highest normalized RMSE, highlighting the benefit of RR-GBM in the reduced rank setting, particularly as the number of sample covariates $L$ grows.}
    \label{fig:sample-covariate-simulation}
\end{figure}

\cref{fig:reduced-rank-performance} illustrates that RR-GBM has better RMSE than the full GBM model when the true $B$ has approximate rank $N_{0} \leq N < L$. 
Indeed, the full GBM corresponds to the model with rank 10 on the x-axis, and we see that the RR-GBM has lower RMSE for every value of $N$ greater or equal to the true approximate rank $N_{0}$.
This result can be intuitively understood as coming from the bias-variance tradeoff, since the increased variance due to estimating the parameters of factors with small effects outweighs the reduction in bias-squared due to having the correct full model. 
 
Even though the true $B$ is full rank, allowing a small amount of bias by ignoring less important factors can reduce the variance in the estimates enough to improve the RMSE.

\cref{fig:sample-covariate-simulation} shows the estimation error as the number of sample covariates $L$ increases, for varying choices of $N$, when $N_0 = 1$. 
As expected, normalized RMSE increases with $L$, however, using an RR-GBM yields much lower error than a full GBM when the true $B$ is approximately reduced rank (\cref{fig:sample-covariate-simulation}). Furthermore, the improvement in relative performance increases with $L$.

\subsection{Computation time of RR-GBM compared to full GBM}

\begin{figure}
    \centering
    \includegraphics[width=0.8\linewidth]{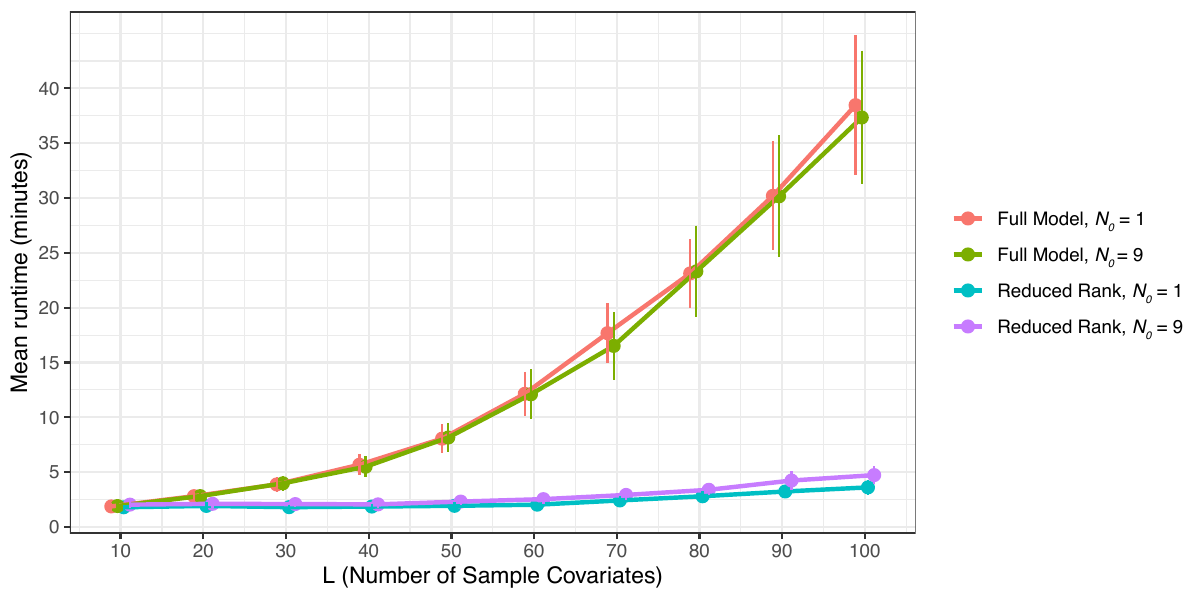}
    \caption{\textbf{Computation time of RR-GBM compared to full GBM.}
    For each $L\in\{10,20,\ldots,100\}$, Poisson simulated data sets were generated using two settings for the approximate rank of the true $B$ matrix: (i) $N_0 = 1$ and (ii) $N_0 = 9$. Two models were fit to each data set: a full-rank model with $N = L$ and a reduced-rank model with $N = N_0$. Each point shown is the average of 100 simulations with the respective settings, along with $I = 5000$, $J = 2500$, $K = 1$, and $M = 1$. Models were run on a single 48-core CPU HPC cluster with at most $15$GB of RAM.}
    \label{fig:speed-comparison}
\end{figure}

In \cref{fig:speed-comparison}, we compare the runtime of the RR-GBM and the full GBM, averaged over 100 simulations.
As the number of sample covariates $L$ increases, the computation time required to fit the full GBM increases rapidly. Meanwhile, the RR-GBM requires only around twice as much time to fit with $L = 100$ sample covariates as with $L = 10$. Also, we see that these computation times do not depend strongly on the true approximate rank $N_{0}$ for $N_{0} \ll L$.  Here, for the RR-GBM, we set the model rank $N$ equal to the true approximate rank $N_0$.

\subsection{Visualization of the sample covariate loadings matrix}
\label{sec:vis-loadings}

The decomposition $Q\Lambda R\T$ yields a method for both visualization of the latent covariate structure and selection of covariates, by plotting the loadings matrix $\Lambda R\T$ as described in \cref{sec:vis-sample-cov}.

We consider two distinct ways of generating the $B$ matrix for simulations. In the first setting, referred to as ``Differential Effects'', we initially generate $B$ as described in \cref{sec:sim-data-generation}, then randomly select a subset of covariates, and for these covariates, we double the entries of $B$ for a random subset of features. Conceptually, in the scRNA-seq setting, this represents the situation where a given sample covariate affects the expression of some percentage of genes. In the second setting, referred to as ``Structured Effects'', we generate $B$ using the procedure described in \cref{sec:sim-data-generation}, except that the $R_{0}$ matrix is generated to have a certain column space, as described in \cref{sec:simulating-sample-covar-effects}.

\begin{figure}
    \centering
    \includegraphics[width=1\linewidth]{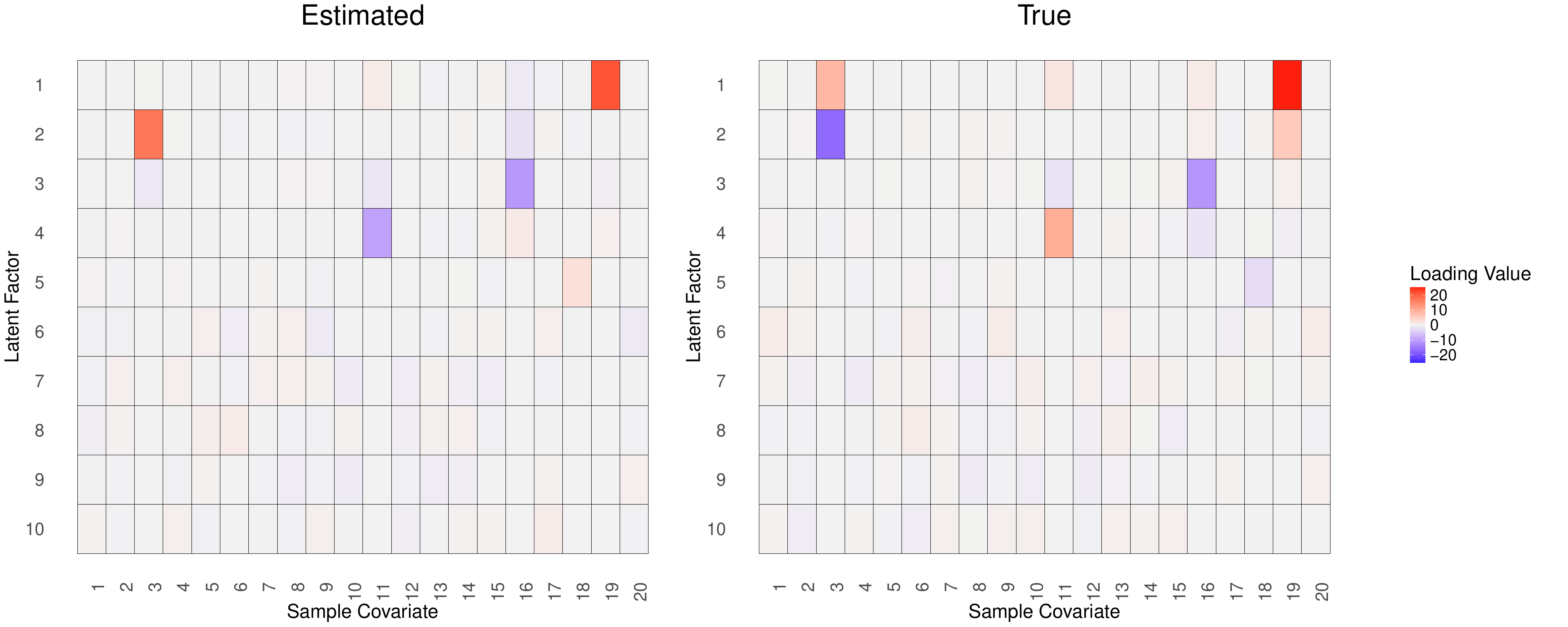}
    \caption{\textbf{Visualizing the effect of sample covariates on latent factors in RR-GBM.}
    The estimated $\hat{\Lambda}\hat{R}\T$ and true $\Lambda_{0}R_{0}\T$ are shown. Data were simulated with 20,000 features, 200 samples, 1 latent factor, and 20 Bernoulli sample covariates with a Negative Binomial outcome distribution. The feature level coefficients for the five sample covariates 18, 11, 16, 3, 19, were altered at random with $\{1\%,\, 5\%,\, 10\%,\, 30\%,\, 50\%\}$ of feature coefficients being doubled, respectively. The true $B_{0}$ matrix was simulated as described in \cref{sec:sim-data-generation}, before the selected coefficients were altered.}
    \label{fig:sample-covariate-visualization}
\end{figure}

First, in the Differential Effects setting, we consider a simulated dataset in which 5 out of 20 covariates affect the outcome, and their effects are uncorrelated.
In \cref{fig:sample-covariate-visualization}, we show the true and estimated loadings. We see that the model correctly infers which covariates are relevant, as well as correctly separating their effects into 5 distinct factors.
 
In this simulation, the magnitude of the loading is related to the percentage of the features that the given covariate influences, with covariate 19 influencing 50\% of features for example. 
Since only the first five factors have non-negligible values in $\Lambda R\T$ for any covariate, the rank of the model can effectively be reduced to five, resulting in a roughly 75\% reduction in the number of parameters that need to be estimated, compared with the full GBM. 

\begin{figure}
    \centering
    \includegraphics[width=1\linewidth]{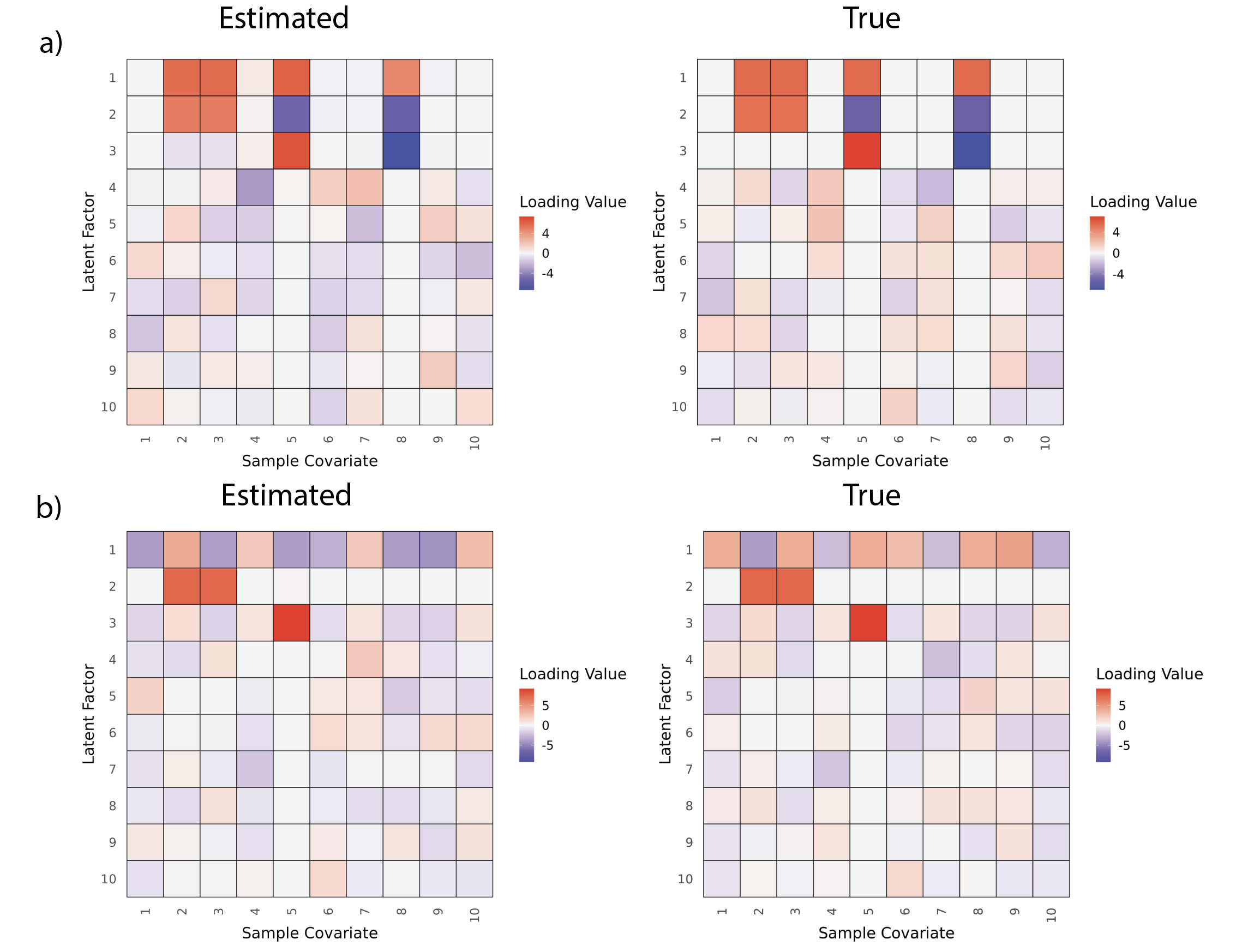}
    \caption{\textbf{Visualizing the effect of sample covariates on latent factors in RR-GBM.} Shown are heatmaps of the estimated and true sample covariate loadings in two scenarios, with the estimated matrix $\hat{\Lambda}\hat{R}\T$ on the left and the ground truth $\Lambda R\T$ on the right. (Scenario a) In the true loadings, four covariates have strong effects on the first three factors, with a nested structure. (Scenario b) In the true loadings, all sample covariates contribute to the first factor, while only covariates two and one contribute to the second and third factors, respectively. 
    In each scenario, $\Lambda_{0}R_{0}\T$ was simulated as described in \cref{sec:paper-figure-struct}.
    Negative Binomial outcome data was simulated with 2000 samples, 5000 features, 10 Bernoulli sample covariates, 1 latent factor, and 1 feature covariate. A full rank 10 model was fit to the data using the RR-GBM estimation algorithm.}
    \label{fig:covar-struct}
\end{figure}

Now, consider the Structured Effects setting.
When the sample covariates affect the outcome in interrelated ways, multiple sample covariates may contribute to a given latent factor. In \cref{fig:covar-struct}, we show estimated RR-GBM loadings for two simulation examples with more complex loadings patterns; see \cref{sec:paper-figure-struct} for details.
These examples demonstrate that the model can distinguish factors with overlapping or even nested sets of sample covariates with nonzero loadings.
Thus, not only is RR-GBM able to identify the relevant covariates, it decomposes their overall effect into loadings onto shared latent factors which, in turn, act in concert on the outcomes.

\subsection{Identification of differential features with respect to each factor}
\label{sec:differential}

Each column of the $Q$ matrix in the decomposition $Q\Lambda R\T$ represents the effect of the corresponding factor on all of the features. 
The $Q$ matrix can be visualized using a heatmap, analogously to  the visualization of the sample covariate loadings matrix $\Lambda R\T$; see \cref{fig:differential-effects-ground-truth-Q} for an example from the Differential Effects simulation.
In particular, this allows for the identification of features that differ with respect to each factor.  With this interpretation, the magnitude of the  entries in the $k$th column of $Q$ indicate the strength of the differential effects of factor $k$ on the respective features. For instance, in scRNA-seq data, a column of $Q$ may represent the effects of a given gene program on the expression levels of all the genes. 

An advantage of the RR-GBM is that it can detect such programs even if the net effect in the $B$ matrix for a given covariate/feature pair is null.  More precisely, suppose $b_{i l} = 0$ for some covariate $l$ and feature $i$. Using a full-rank GBM would reveal no relationship between $l$ and $i$.  However, since the factors are shared among all covariates in an RR-GBM, it is possible to infer nonzero contributions $q_{i n}$ and $r_{l n}$ in the decomposition $b_{i l} = \sum_{n=1}^N \lambda_{n n} q_{i n} r_{l n}$.  Thus, the RR-GBM can yield insights into the effect of covariate $l$ on programs that modulate feature $i$, even when they cancel out in the aggregate for that covariate. This capability is due to the information sharing implied by the reduced-rank structure.

To recover the usual log-fold change estimates for a given feature and sample covariate pair, we can take the product $\hat{Q}\hat{\Lambda}\hat{R}\T$ to obtain $\hat{B}$ and examine the corresponding entry $\hat{b}_{i l}$. 

\subsection{Empirical rank selection}

As shown above, when the true $B$ matrix is exactly or approximately reduced rank, fitting a reduced-rank model results in improved estimates of $B$ compared to using a full GBM. In practice, the rank or approximate rank of $B$ is rarely known and thus it becomes necessary to have a selection method for choosing the rank of $B$ in the fitted model. It is not possible to use a standard cross-validation approach since sample- and gene-specific parameters are fit in the model. To overcome this obstacle, we use the data thinning technique of \citet{neufeld_inference_2024} to create a train and test split of our data and assess prediction performance via RMSE. This procedure creates two independent data matrices, each having the same dimensions as the original data matrix but with half the mean. Thus, we can use one to fit and one to assess performance. The rank of $B$ that yields the smallest test RMSE then provides a data-driven choice of rank to use in the model. 

\begin{figure}
    \centering
    \includegraphics[width=1\linewidth]{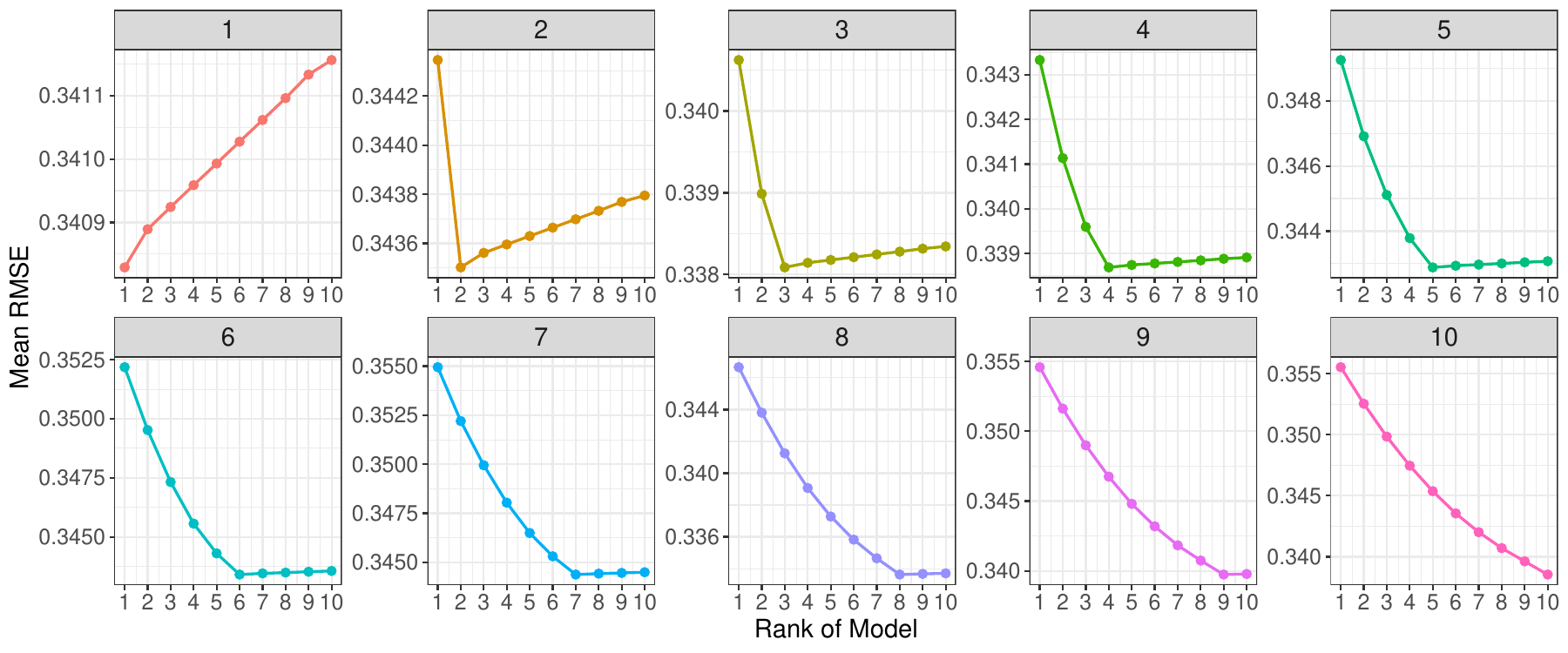}
    \caption{\textbf{Empirical selection of the rank in RR-GBM.}
    Panels are labeled according to the  true approximate rank of $B$, where the corresponding $L-N_{0}$ singular values are 0.01 times their randomly generated value to create a $B$ with approximate rank $N_{0}$. Each dot indicates the average RMSE (over 100 simulations) between the predicted output values and the true values. Poisson data was generated and fit using a model with 1 latent factor ($M =1$).}
    \label{fig:empirical-rank-selection}
\end{figure}

\cref{fig:empirical-rank-selection} demonstrates this rank selection procedure on data sets generated using true $B$ matrices with approximate ranks ranging from $1$ to $10$; see \cref{sec:sim-data-generation} for details.  In each case, we find that the mean test RMSE from 100 simulations is minimized at the true approximate rank of $B$. This same pattern was observed in all 100 simulations and a version of this figure from one simulation run can be seen in \cref{fig:empirical-rank-selection-single-run}. 

In addition to selection of rank of $B$ in the reduced-rank regression setting, this method can also be used to select $M$, the rank of $U \Sigma V\T$, in both the RR-GBM and the full GBM. By performing the same data thinning technique and fitting across a range of $M$ values, this procedure provides an estimate of the number of latent factors in the true data generating process.

\section{Application}
\label{sec:application}

The RR-GBM model lends itself to applications in which there are a large number of sample covariates that are potentially important to consider.  In particular, this is the case when the primary objective is to determine which of many sample covariates are meaningful and what their effects are. 
An excellent example is single-cell RNA-sequencing screening, an experimental protocol that has recently been growing in popularity.

In an scRNA-seq screen, a range of perturbations are performed, such as genetic or environmental modifications, and scRNA-seq is used to quantify the resulting transcriptomic profiles. Typically, the experiment is designed such that each cell is perturbed by deleting, amplifying, or inhibiting the activity of one or more genes or exposing the cell to an environmental agent, such as a drug or biologically active compound. 

In the gene perturbation setting, the number of genes that are perturbed can range from hundreds to the entire collection of 20,000 genes, and in the environmental perturbation setting, there is no limit to the number of perturbations that can be performed.
The objective is to infer which perturbations have significant effects, and how they affect the expression of genes. 
Thus, from a statistical perspective, each different perturbation represents a binary sample covariate that equals 0 or 1, indicating whether that perturbation was applied to the given cell.

We refer to \citet{adamson_multiplexed_2016} and \citet{dixit_perturb-seq_2016} for background.

We apply RR-GBM to an scRNA-seq screen data set consisting of the transcriptional responses of pancreatic cancer cells perturbed by a range of tumor microenvironment (TME) protein ligands \citep{liu_scalable_2025}. The data consist of counts for $I = 15{,}876$ genes in $J = 10{,}881$ cells exposed to $L = 68$ different ligands; see Liu et al. 2025 for experimental design details. 

The original study of \citet{liu_scalable_2025} analyzed the data using a pipeline in which they first run consensus non-negative matrix factorization (cNMF) to identify gene expression programs (GEPs), filter to GEPs that were variably expressed across cells, and run elastic net regression to infer the effect of each perturbation on the GEPs.
For visualization, they compute Pearson residuals, run principal components analysis, use the elbow method to select the number of components, and then run UMAP.
In contrast, we simply fit an RR-GBM to the raw count data without any filtering.
The columns of $Q$ in the RR-GBM can be interpreted as orthogonal GEPs, aiding the biological interpretation, and the effects of the perturbations on the GEPs are captured by the loadings $\Lambda R\T$. Thus, the RR-GBM provides a more direct analytical approach using one coherent statistical model.

\begin{figure}
    \centering
    \includegraphics[width=1\linewidth]{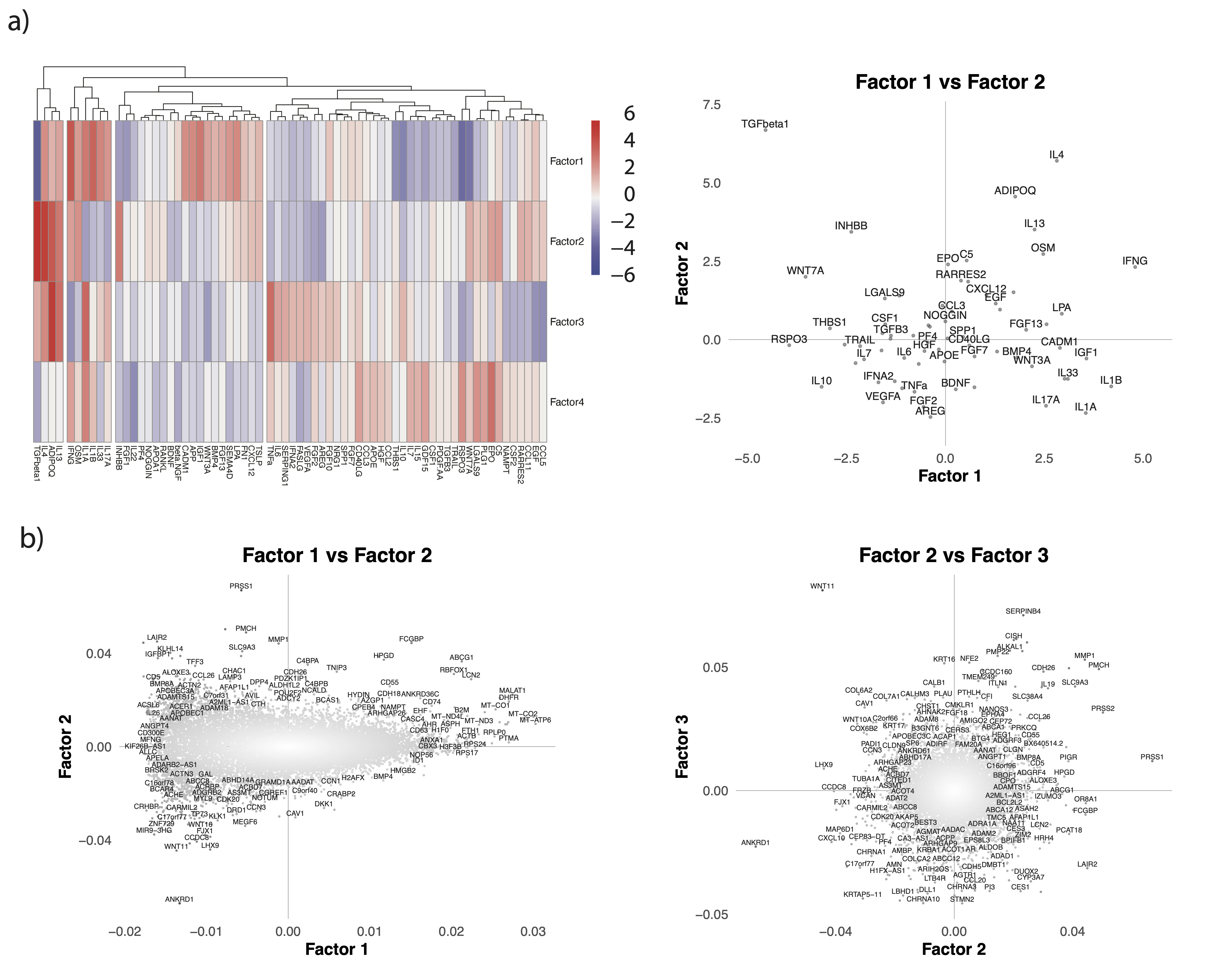}
    \caption{\textbf{RR-GBM results on Perturb-seq data for TME ligand experiment.}
    We fit an RR-GBM with rank $N=4$ to the \citep{liu_scalable_2025} dataset of a TME phenotypic screen in PDAC organoids. (a) $\hat{\Lambda} \hat{R}\T$ is shown on the left, with hierarchical clustering is applied to the TME ligands (columns). The dendogram was cut into 4 groups to highlight a potential grouping of ligands. The panel on the right shows an alternate visualization of the top two factors in $\hat{\Lambda} \hat{R}\T$, with factor 2 plotted against factor 1. (b) The figure on the left shows the first and second columns of $\hat{Q}$ plotted against each other to highlight differential genes between them. The figure on the right shows the second and third columns of $\hat{Q}$ plotted against each other.}
    \label{fig:phenotypic-screen}
\end{figure}

\cref{fig:phenotypic-screen} summarizes the results using a Negative Binomial RR-GBM with rank $N=4$, which yields many of the same results as \citet{liu_scalable_2025} while using a more unified approach.
In \cref{fig:phenotypic-screen}a, we visualize the relationships among the 68 perturbations via a heatmap of $\hat{\Lambda}\hat{R}\T$ as described in \cref{sec:vis-sample-cov}, with the addition of column (ligand) ordering by hierarchical clustering over the 4 latent factors.
The ligands with large magnitude loadings in factor~1, such as IFNG, IL1A, and IL1B, are related to inflammatory and growth related signaling; meanwhile, ligands with large magnitude loadings in factor 2, such as TGF$\beta$, ADIPOQ, and IL4, are related to immune regulation and tissue remodeling \citep{lee_heterocellular_2021, wynn_type_2015, liu_scalable_2025}.

\begin{figure}
    \centering
    \includegraphics[width=0.9\linewidth]{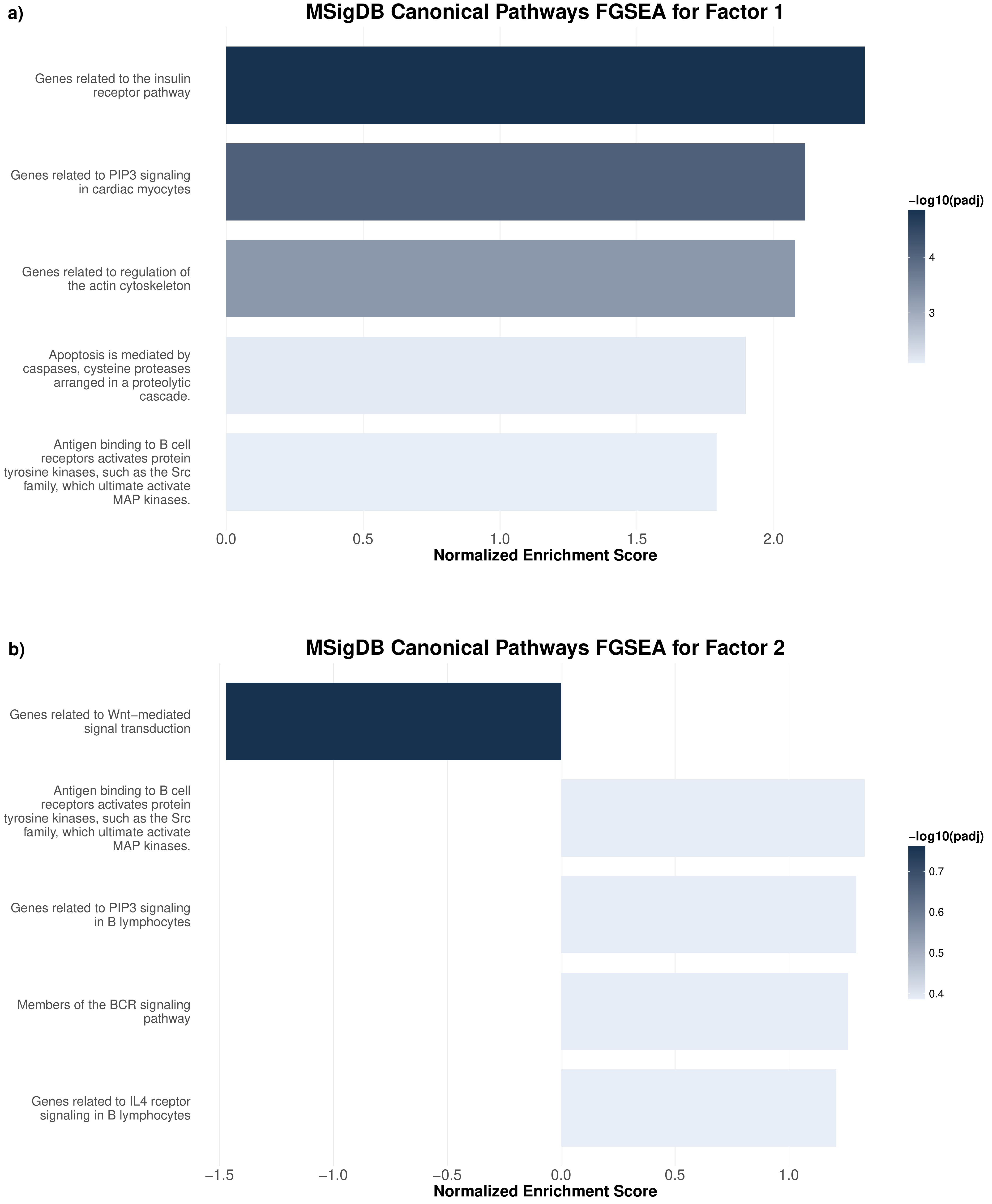}
    \caption{\textbf{FGSEA of gene loadings for factors 1 and 2.} The columns of $\hat{Q}$ were used as rankings for the FGSEA algorithm \citep{korotkevich2016fast} to identify relevant pathways that are affected by the corresponding identified ligands. Pathways from the canonical pathways gene set collection from MSigDB were used \citep{subramanian_gene_2005, liberzon2011molecular}. (a) The canonical pathways most associated with the gene loadings from factor 1. (b) The canonical pathways most associated with the gene loadings from factor 2.}
    \label{fig:fgsea}
\end{figure}

To investigate the effects of these two factors on gene expression, we visualize the corresponding columns of $\hat{Q}$ in a scatterplot (\cref{fig:phenotypic-screen}b). Since the sign of each factor is arbitrary, the positive/negative direction of each axis in \cref{fig:phenotypic-screen}b needs to be interpreted in conjunction with the corresponding row in \cref{fig:phenotypic-screen}a. For example, we see in \cref{fig:phenotypic-screen}a that the second cluster of ligands---consisting of IL17A, IL33, IL1B, IL1A, OSM, and IFNG---all have positive loadings for factor 1, which means that genes with a positive factor 1 value in \cref{fig:phenotypic-screen}b are upregulated in cells exposed to these ligands. To link the results with known biology, we can use standard methods such as enrichment analysis to gain insight into pathways associated with each inferred GEP \citep{korotkevich2016fast}. For instance, in \cref{fig:fgsea}, we find that notable pathways associated with factor 1 include regulation of the cytoskeleton and PI3K signaling,  and notable pathways with factor 2 include down-regulation of Wnt signaling and immune mediated signaling. While experimental validation of these findings are outside the scope of this work, we refer to \citet{liu_scalable_2025}, where findings related to the PDAC IL-4/IL-13 response program are captured in our factor 1 and the TGF$\beta$ program is similar to our factor 2.  
Taken together, these results illustrate how the estimated $\hat{Q}$, $\hat{\Lambda}$, and $\hat{R}$ 
can be used directly for visualization, analysis, and hypothesis generation in a unified statistical framework without the need for additional pre- or post-processing.

\section{Discussion}
\label{sec:discussion}

The RR-GBM makes it possible to apply the GBM framework to data matrices with a large number of sample covariates.
Specifically, by constraining the sample covariate parameter matrix $B$ to be low rank, the RR-GBM yields improved estimates at lower computational cost.
The resulting factorization of $B$ enables visualization of the relationships among sample covariates in terms of their effects, and identification of biologically meaningful latent factors such as gene expression programs. Our data thinning approach for selecting the best-performing rank of $B$ to use in the RR-GBM is easily conducted in parallel and does not impose any additional computational costs beyond fitting the model a selected number of times.

In future work, the proposed method could also be applied to the feature covariate matrix $A$, to handle large numbers of feature covariates; by the symmetry of \cref{eq:gbm}, the approach we developed for $B$ can alternatively be used to estimate a reduced-rank $A$ matrix.
Additionally, providing uncertainty quantification for the entries of the reduced-rank matrix components would enable confidence interval construction and hypothesis testing on both the latent factor directions ($Q$) and the associated loadings onto the sample covariates ($R$). This capability would be useful for identifying statistically significant genes in a gene expression program or detecting differentially expressed genes between two groups.
Another interesting direction would be to apply the iteratively reweighted singular value decomposition (IRSVD) algorithm of \citet{nicol_model-based_2025} to estimate $Q\Lambda R\T$ instead of the bounded regularized Fisher scoring approach used in the present paper, in order to scale up to even larger numbers of features and sample covariates.

\section*{Acknowledgments}

We thank Rafael Irizarry and Rong Ma for helpful discussions. This work was funded in part by the NIH Genomics Training Grant (T32GM135117). The content is solely the responsibility of the authors and does not necessarily
represent the official views of the National Institutes of Health.

\bibliographystyle{chicago}
\bibliography{references}


\clearpage
\setcounter{page}{1}
\setcounter{section}{0}
\setcounter{table}{0}
\setcounter{figure}{0}
\renewcommand{\theHsection}{SIsection.\arabic{section}}
\renewcommand{\theHtable}{SItable.\arabic{table}}
\renewcommand{\theHfigure}{SIfigure.\arabic{figure}}
\renewcommand{\thepage}{S\arabic{page}}  
\renewcommand{\thesection}{S\arabic{section}}   
\renewcommand{\thetable}{S\arabic{table}}   
\renewcommand{\thefigure}{S\arabic{figure}}

\begin{center}
{\Large Supplementary material for ``Reduced-rank generalized bilinear models''}
\end{center}

\thispagestyle{empty}

\vspace{1em}

\section{Detailed algorithm for RR-GBM estimation}
\label{sec:estimation-algo}

In this section, we describe our algorithm for estimating the $Q$, $\Lambda$, and $R$ parameter matrices.
For details on estimating the rest of the GBM parameters, we refer to \citet{miller_inference_2020}. The following steps are substituted into the algorithm of \citet{miller_inference_2020} in place of the $B$ estimation steps. We cycle through the steps to update each parameter matrix, iterating until convergence as in \citet{miller_inference_2020}.

We first define non-standard notation, following the conventions used in \citet{miller_inference_2020}. We write $\mathrm{Block}(A_{ij}: i,j \in \{1, \dots, n\})$ to denote the block matrix with block $i, j$ equal to the matrix $A_{ij}$. For a matrix $A \in \R^{m\times n}$, $A_{i*}$ and $A_{*j}$ denote the diagonal matrices constructed from the $i$th row and the $j$th column, respectively. For $A \in \R^{m\times n}$, $\mathrm{vec}(A)$ is defined as $\mathrm{vec}(A) = (a_{11}, a_{21}, \dots, a_{m1}, ,a_{12}, a_{22}, \dots, a_{m2}, \dots, a_{mn}) \in \R^{mn}$. ``Unvectorize'' returns the original matrix $A$ that has been vectorized as implemented by $\mathrm{vec}(A)$. $X^{+}$ is the Moore--Penrose pseudoinverse of $X$.

The steps for updating $Q$, $\Lambda$, and $R$ are as follows. The matrices $W$ and $E$ are defined as in Section~S5 in \citet{miller_inference_2020}.

\begin{itemize}[itemsep=0pt, topsep=0pt, parsep=0pt, partopsep=0pt]
    \item Updating $Q\Lambda$
    \begin{enumerate}
        \item $\Phi \leftarrow Q\Lambda$
        \item For $i = 1, \ldots, I$
        \begin{enumerate}
            \item $\xi \leftarrow (R\T Z\T W_{i*}ZR + \lambda_{b}I)^{-1}(R\T Z\T E_{i:} - \lambda_{b}\Phi_{i:})$
            \item $\Phi_{i:} \leftarrow \Phi_{i:} + \xi \min\{1, \rho\sqrt{\mathrm{dim}(\Lambda)}/\norm{\xi}\}$
        \end{enumerate}
        \item $B \leftarrow \Phi R\T$
        \item $D \leftarrow X^{+}B$
        \item $B \leftarrow B - XD$
        \item $C \leftarrow C + D$
        \item Run compact SVD of rank $N$ on $B$ yielding $Q\Lambda R\T$
    \end{enumerate}
    \item Updating $\Lambda R\T$
    \begin{enumerate}
        \item $O \leftarrow \Lambda R\T$
        \item $F \leftarrow \mathrm{Block}(\sum_{j=1}^{J}z_{j\ell} z_{j\ell'}(Q\T W_{*j}Q): \ell, \ell' \in \{1, \ldots, L\})$
        \item $\xi\leftarrow (F + \lambda_{b}I)^{-1}(\mathrm{vec}(Q\T EZ) - \lambda_{b} - \lambda_{b}\mathrm{vec}(O))$
        \item $\mathrm{vec}(O) \leftarrow \mathrm{vec}(O) + \xi\min\{1, \rho\sqrt{\mathrm{dim}(\Lambda)}/\norm{\xi}\}$
        \item Unvectorize $O$
        \item $B \leftarrow \Rho O$
        \item $D \leftarrow X^{+}B$
        \item $B \leftarrow B - XD$
        \item $C \leftarrow C + D$
        \item Run compact SVD of rank $N$ on $B$ yielding $Q\Lambda R\T$
    \end{enumerate}
    \item Updating $\Lambda$
    \begin{enumerate}
        \item $F \leftarrow \sum_{i=1}^{I}(ZRQ_{i*})\T W_{i*}ZRQ_{i*}$
        \item $\xi \leftarrow (F + \lambda_{b}I)^{-1}(\mathrm{diag}(Q\T E Z R)-\lambda_{b}\mathrm{diag}(\Lambda))$  
        \item $\mathrm{diag}(\Lambda) \leftarrow \mathrm{diag}(\Lambda) + \xi\min\{1, \rho\sqrt{\mathrm{dim}(\Lambda)}/\norm{\xi}\}$
    \end{enumerate}
\end{itemize}

\section{Identifiability constraints for RR-GBM}
\label{sec:complete-ident-constraints}

Recall the form of the RR-GBM model in \cref{eq:rrgbm}:
$$
g(\E Y) = X A\T + Q \Lambda R\T Z\T + X C Z\T + U \Sigma V\T.
$$
We show that the following constraints ensure identifiability of all the model parameters:

\begin{enumerate}[label=(\alph*)]
    \item\label{item:identifiability1} $X\T X$ and $Z\T Z$ are invertible,
    \item $X\T Q\Lambda R\T = 0$, $Z\T A = 0$, $X\T U = 0$, and $Z\T V = 0$,
    \item\label{item:identifiability3} $U\T U = \mathrm{I}$, $V\T V = \mathrm{I}$, $Q\T Q = \mathrm{I}$, and $R\T R = \mathrm{I}$,
    \item $\Sigma$ is a diagonal matrix such that $\sigma_{11} > \sigma_{22} > \cdots > \sigma_{MM} > 0$,
    \item the first nonzero entry of each column of $U$ is positive,
    \item\label{item:identifiability6} $\Lambda$ is a diagonal matrix such that $\lambda_{11} > \lambda_{22} > \cdots > \lambda_{NN} > 0$, and
    \item\label{item:identifiability7} the first nonzero entry of each column of $Q$ is positive.
\end{enumerate}

where $A \in \R^{J\times K}$, $C \in\R^{K\times L}$, $\Sigma\in\R^{M\times M}$, $U \in\R^{I\times M}$, $V\in\R^{J\times M}$, $X\in\R^{I\times K}$, $Z\in\R^{J\times L}$, $Q \in \mathbb{R}^{I \times N}$, $\Lambda \in \mathbb{R}^{N \times N}$, $R \in\mathbb{R}^{L \times N}$, $N < \min\{I, L\}$, and $ M < \min\{I, J\}$. 

Since $B = Q\Lambda R\T$, the proof of identifiability for $A$, $C$, $U$, $\Sigma$, and $V$ is still valid and can be found in \citet{miller_inference_2020}.
We now show that $Q$, $\Lambda$, and $R$ are also identifiable given these constraints.
If $(A_{1}, Q_{1}, \Lambda_{1}, R_{1}, C_{1}, U_{1}, \Sigma_{1}, V_{1}, X, Z)$ and $(A_{2}, Q_{2}, \Lambda_{2}, R_{2}, C_{2}, U_{2}, \Sigma_{2}, V_{2}, X, Z)$ satisfy the above constraints and 
\begin{align}
\label{eq:identifiability}
X A_{1}\T + Q_{1}\Lambda_{1}R_{1}\T Z\T + X C_{1}Z\T + U_{1}\Sigma_{1}V_{1}\T = X A_{2}\T + Q_{2}\Lambda_{2}R_{2}\T Z\T + X C_{2}Z\T + U_{2}\Sigma_{2}V_{2}\T 
\end{align}
then we know from \citet{miller_inference_2020} that $A_{1} = A_{2}$, $C_{1} = C_{2}$, $U_{1} = U_{2}$, $\Sigma_{1} = \Sigma_{2}$, and $V_{1} = V_{2}$. Plugging these equations into \cref{eq:identifiability} and canceling yields that
\[Q_{1}\Lambda_{1}R_{1}\T Z\T = Q_{2}\Lambda_{2}R_{2}\T Z\T.\]
Right multiplying by $Z$ and using constraint \ref{item:identifiability1}, we have

\[Q_{1}\Lambda_{1}R_{1}\T = Q_{2}\Lambda_{2}R_{2}\T.\]

By the uniqueness properties of the singular value decomposition, constraints \ref{item:identifiability3} and \ref{item:identifiability6} imply that $\Lambda_{1} = \Lambda_{2}$, $Q_{1} = Q_{2}S$, and $R_{1}\T = SR_{2}\T$ for a diagonal matrix $S$ of the form $S = \mathrm{Diag}(\pm1, \dots, \pm1)$ \citep{blum_foundations_2020}. By constraint \ref{item:identifiability7}, $S = \mathrm{I}$, and hence, $Q_{1} = Q_{2}$ and $R_{1} = R_{2}$. Therefore, by \cref{eq:rrgbm}, $Q$, $\Lambda$, and $R$ are uniquely determined by $\E Y$ for any given $X$, $Z$, $M$, and $N$.

\section{Parameter count for $B$ in the GBM and RR-GBM}
\label{sec:number-of-parameters}

In this section, we count the number of free parameters in the $B$ matrix under the assumed identifiability constraints in \cref{sec:complete-ident-constraints}. We consider both the case of the full GBM and the RR-GBM.

\subsection{Full GBM setting}
\label{sec:number-of-parameters-GBM}

In the full GBM, the only constraint on $B$ is $X\T B = 0$. Since $X \in \R^{I\times K}$ and $B \in \R^{I \times L}$, the constraint $X\T B = 0$ is equivalent to the requirement that $X\T b_{l} = 0$ for $l=1,\ldots,L$, where $b_{1},\ldots,b_{L}$ are the columns of $B$. The $K$ columns of $X$ are linearly independent since $X\T X$ is invertible by assumption. Thus, the space of solutions to $X\T b_{l} = 0$ has dimension $I - K$ by the rank-nullity theorem; in other words, $b_{l}$ has $I - K$ free parameters. Thus, over all $L$ columns of $B$, the total number of free parameters to be estimated, given the constraint $X\T B = 0$, is $(I-K)L$.

\subsection{RR-GBM setting}
\label{sec:number-of-parameters-RRGBM}

In the RR-GBM, $B$ is parametrized as $B = Q\Lambda R\T$, where $Q \in \R^{I\times N}$, $\Lambda \in \R^{N \times N}$, $R \in \R^{L\times N}$, and $\Lambda$ is diagonal with non-zero entries on the diagonal, and from \cref{sec:complete-ident-constraints}, the assumed identifiability constraints are: 

\begin{enumerate}
    \item $Q\T Q = \mathrm{I}$,
    \item $R\T R = \mathrm{I}$,
    \item $\lambda_{11} > \lambda_{22} > \cdots > \lambda_{NN} > 0$, and
    \item $X\T Q\Lambda R\T = 0$.
\end{enumerate}

We also assume $N < \min\{L, I\}$ and $K < I$, which are satisfied in all our applications of the model.
To count the number of free parameters in $B$ under these constraints, we first argue that given constraints 1-3, the fourth constraint ($X\T Q\Lambda R\T = 0$) is equivalent to $X\T Q = 0$.  To see this, let $S = X\T Q \Lambda$  and note that the constraint becomes $S R\T = 0$. Since $R\T R = \mathrm{I}$, the columns of $R$ are linearly independent, and thus its null space is $\{0\}$. Hence, $SR\T = 0$ implies $S = 0$ since the left null space of $R^{T}$ is same as the null space of $R$. Thus, since $R\T R = \mathrm{I}$, the constraint $X\T Q\Lambda R\T = 0$ holds if and only if $X\T Q\Lambda = 0$.
Furthermore, since $\Lambda$ is a diagonal matrix of non-zero entries,  $X\T Q\Lambda = 0$ if and only if $X\T Q = 0$. Therefore the constraints can equivalently be written:

\begin{enumerate}
    \item $Q\T Q = \mathrm{I}$,
    \item $R\T R = \mathrm{I}$,
    \item $\lambda_{11} > \lambda_{22} > \cdots > \lambda_{NN} > 0$, and
    \item $X\T Q = 0$.
\end{enumerate}

To count the free parameters, begin by noting that $\Lambda$ has $N$ free parameters. Next, the constraint that $R\T R = \mathrm{I}$ is equivalent to $R \in V_{N}(\R^{L})$, where $V_{N}(\R^{L})$ denotes the Stiefel manifold represented by the set of all orthonormal $N$-frames in $\R^{L}$. Thus, $R$ has $L N - N(N+1)/2$ free parameters \citep{chakraborty_statistics_2019}.
Finally, $Q$ is subject to the constraints that $Q\T Q = \mathrm{I}$ and $X\T Q = 0$. 
The $X\T Q = 0$ constraint forces $Q$ to lie in the null space of $X\T$, which has dimension $I-K$ since $X$ is rank $K$.
Thus, these constraints force $Q$ to be an orthonormal $N$-frame in an $I-K$ dimensional subspace, which is equivalent to $V_{N}(\R^{I-K})$, and this Stiefel manifold has dimension $(I-K)N - N(N+1)/2$. Hence, $Q$ has $(I-K)N - N(N+1)/2$ free parameters.

Summing up the number of free parameters for $\Lambda$, $R$, and $Q$ yields a total of
$$ N + L N - \frac{N(N+1)}{2} + (I-K) N - \frac{N(N+1)}{2}, $$
which simplifies to $N(I-K+L-N)$.

\subsection{Reduction in parameter count for RR-GBM relative to full GBM}
\label{subsubsec:reduc-in-param}

Let $p_\textsc{gbm}$ and $p_\textsc{rrgbm}$ denote the number of parameters in $B$ under the GBM and the RR-GBM, respectively. 
By \cref{sec:number-of-parameters-GBM,sec:number-of-parameters-RRGBM}, the fraction of parameters under the RR-GBM relative to the GBM is 
$$ \frac{p_\textsc{rrgbm}}{p_{\textsc{gbm}}} = \frac{N(I-K+L-N)}{(I-K)L} = \frac{N}{L} + \frac{N}{I-K} - \frac{N}{L}\frac{N}{I-K}. $$

Meanwhile, the reduction in the number of parameters, represented as a fraction, is
$$ 1 - \frac{p_\textsc{rrgbm}}{p_{\textsc{gbm}}} = 1 - \frac{N}{L} - \frac{N}{I-K} + \frac{N}{L}\frac{N}{I-K} = \bigg(1 - \frac{N}{I-K}\bigg)\bigg(1 - \frac{N}{L}\bigg).$$
This expression highlights the reliance on both the ratio of $N$ to $L$ and also the ratio of $N$ to $I-K$.
In particular, if $I - K\gg N$, then 
$$ \frac{p_\textsc{rrgbm}}{p_{\textsc{gbm}}} \approx \frac{N}{L} \qquad\text{and}\qquad 1 - \frac{p_\textsc{rrgbm}}{p_{\textsc{gbm}}} \approx 1-\frac{N}{L}. $$

\section{Details of the simulation studies}
\label{sec:sim-data-generation}

In this section, we provide additional details on the simulation studies in \cref{sec:simulations}.

\subsection{Simulation of covariates, true parameters, and data}

The true parameters $A_{0}$, $C_{0}$, $U_{0}$, $\Sigma_{0}$, $V_{0}$, $S_{0}$, $T_{0}$, and $\omega_{0}$ as well as the covariate matrices $X$ and $Z$ are generated using the same procedure as described by section S2 in \citet{miller_inference_2020}, where $S_{0}$, $T_{0}$, and $\omega_{0}$ are only used in the Negative Binomial setting. The additional RR-GBM true parameters $Q_{0}\in\R^{I\times N}$, $\Lambda_{0}\in\R^{N\times N}$, and $R_{0}\in\R^{L\times N}$ are simulated as follows, assuming $I > L$. 
We generate $Q_{0}$ and $R_{0}$ uniformly from the Stiefel manifolds $V_{N}(\R^{I})$ and $V_{N}(\R^{L})$, respectively. 
For instance, in the case of $Q_{0}$, this is done by first generating $\tilde{Q}\in\R^{I\times N}$ with i.i.d.\ entries $\tilde{q}_{i n} \sim \mathcal{N}(0, 1)$, and then computing the thin QR decomposition of $\tilde{Q}$ to obtain $\tilde{Q} = Q_{0}Q_\mathrm{R}$ where $Q_0\in\R^{I\times N}$ and $Q_\mathrm{R}\in\R^{N\times N}$. It can be shown that this $Q_{0}$ is uniformly sampled from the Stiefel manifold $V_{N}(\R^{I})$ \citep{tropp_comparison_2012}. The diagonal entries of $\Lambda_{0}$ are chosen to be evenly spaced from $\sqrt{I} + \sqrt{L}$ to $2(\sqrt{I} + \sqrt{L})$, following the approach of \citet{miller_inference_2020}, which is based on the distribution of singular values in random matrices \citep{marcenko_distribution_1967}. To make the $B_{0}$ matrix have approximate rank $N$ (but exact rank $L$), we multiply the last $L - N$ diagonal entries by a small scaling factor equal to $0.01$ by default. (To make $B_0$ exactly reduced-rank, we use a multiplier of $0$ here.) We then compute $\tilde{B} = Q_{0}\Lambda_{0}R_{0}\T$ and define the true $B$ matrix to be $B_{0} = \tilde{B} - X(X^{+}\tilde{B})$, where $X^{+}$ is the Moore--Penrose pseudoinverse of $X$.  This last projection step ensures that $B_0$ satisfies the constraint that $X\T B_0 = 0$, while preserving the reduced-rank or approximate reduced-rank property. We then apply the SVD to $B_{0}$ once more to get the final true values of $Q_{0}$,$\Lambda_{0}$, and $R_{0}$.

Finally, the data matrix $Y$ is then generated in the same way as described in section S2 in\citet{miller_inference_2020}, using the $\log$ link function and either the Poisson or Negative Binomial distribution for sampling as indicated.

\subsection{Structure in the sample covariate effects matrix}
\label{sec:simulating-sample-covar-effects}

Given a set of sample covariates $Z$, we can simulate any desired latent sample covariate effect structure through the controlled generation of $R_{0}$, enabling the construction of an $R_{0}$ with any desired column space. 
The different sample covariate effect structures 
were all simulated with the same underlying hyperparameters: $I = 5000$, $J = 2000$, $K = 1$, $L = 10$, and $M = 1$.  
The $R_{0}$ matrix was simulated to meet the identifiability constraints with the desired structure, denoted by $\tilde{R}$, a matrix of dimension $L \times P$ where $P$ is the desired number of factors with $P \leq N$. It is assumed that the columns of $\tilde{R}$ are orthogonal, as if they were not the final $R_{0}$ would not lie in the Stiefel manifold. An example of $\tilde{R}$ could be a binary matrix encoding a desired sparsity pattern such as that found in \cref{fig:sample-covariate-visualization}. 
The sampling process can be described as uniform sampling from the Stiefel manifold with the respective dimensions subject to the constraint that the column normalized $\tilde{R}$ is a submatrix of the final $R_{0}$. 
 The steps of the sampling procedure are as follows:

\begin{enumerate}
    \item Set factor $\tilde{R} \in \R^{L \times P}$ to have the desired latent sample covariate structure.
    \item $R_{*j} = \tilde{R}_{*j}/\lVert \tilde{R}_{*j}\rVert$ (column-wise normalization)
    \item Simulate $G \in \R^{L \times (N-P)}$ where $G_{ij} \sim \mathcal{N}(0,1)$.
    \item $H = (\textbf{I} - RR\T)G$
    \item Take QR decomposition of $H$ to obtain $H_\mathrm{Q}$ where $H = H_\mathrm{Q}H_\mathrm{R}$.
    \item Create the simulated true $R_{0}$ matrix via the following construction $R_{0} = \begin{bmatrix} R &H_\mathrm{Q}\end{bmatrix}$ (or placing $R$ in the desired column locations).
\end{enumerate}

The resulting $R_{0}$ matrix has the constructed effect on the sample covariate effects as represented by $\tilde{R}$ since the remaining $H_{Q}$ columns are the result of projecting Gaussian noise onto the orthogonal complement of the space spanned by $\tilde{R}$.

For the situation when $\tilde{r} = \tilde{R}$ is a vector, the same procedure holds and generates the desired $R_{0}$.

\subsubsection{Structure in sample covariate effect matrices used}
\label{sec:paper-figure-struct}

\cref{fig:covar-struct}(a) was generated using the following $\tilde{R}$:

\[
\tilde{R} = \begin{bmatrix}
0 &  0 &  0 \\
1 &  1 &  0 \\
1 &  1 &  0 \\
0 &  0 &  0 \\
1 & -1 &  1 \\
0 &  0 &  0 \\
0 &  0 &  0 \\
1 & -1 & -1 \\
0 &  0 &  0 \\
0 &  0 &  0
\end{bmatrix}.
\]
  
\cref{fig:covar-struct}(b) was generated using the following $\tilde{R}$; here, column 3 was obtained by applying Gram-Schmidt orthogonalization to the matrix $\tilde{R}_0$ shown on the left:
\[
\begin{aligned}
\tilde{R}_0 &=
\begin{bmatrix}
 0.8 & 0 & 0 \\
-0.8 & 1 & 0 \\
 0.8 & 1 & 0 \\
-0.5 & 0 & 0 \\
 0.8 & 0 & 1 \\
 0.6 & 0 & 0 \\
-0.5 & 0 & 0 \\
 0.8 & 0 & 0 \\
 0.9 & 0 & 0 \\
-0.6 & 0 & 0
\end{bmatrix},
\qquad
&
\tilde{R} &=
\begin{bmatrix}
 0.8 & 0 & -0.13 \\
-0.8 & 1 & 0.13 \\
 0.8 & 1 & -0.13 \\
-0.5 & 0 & 0.08 \\
 0.8 & 0 & 0.94 \\
 0.6 & 0 & -0.10 \\
-0.5 & 0 & 0.08 \\
 0.8 & 0 & -0.13 \\
 0.9 & 0 & -0.15 \\
-0.6 & 0 & 0.10
\end{bmatrix}.
\end{aligned}
\]

\clearpage
\section{Supplemental figures}

\begin{figure}[H]
    \centering
    \includegraphics[width=1\linewidth]{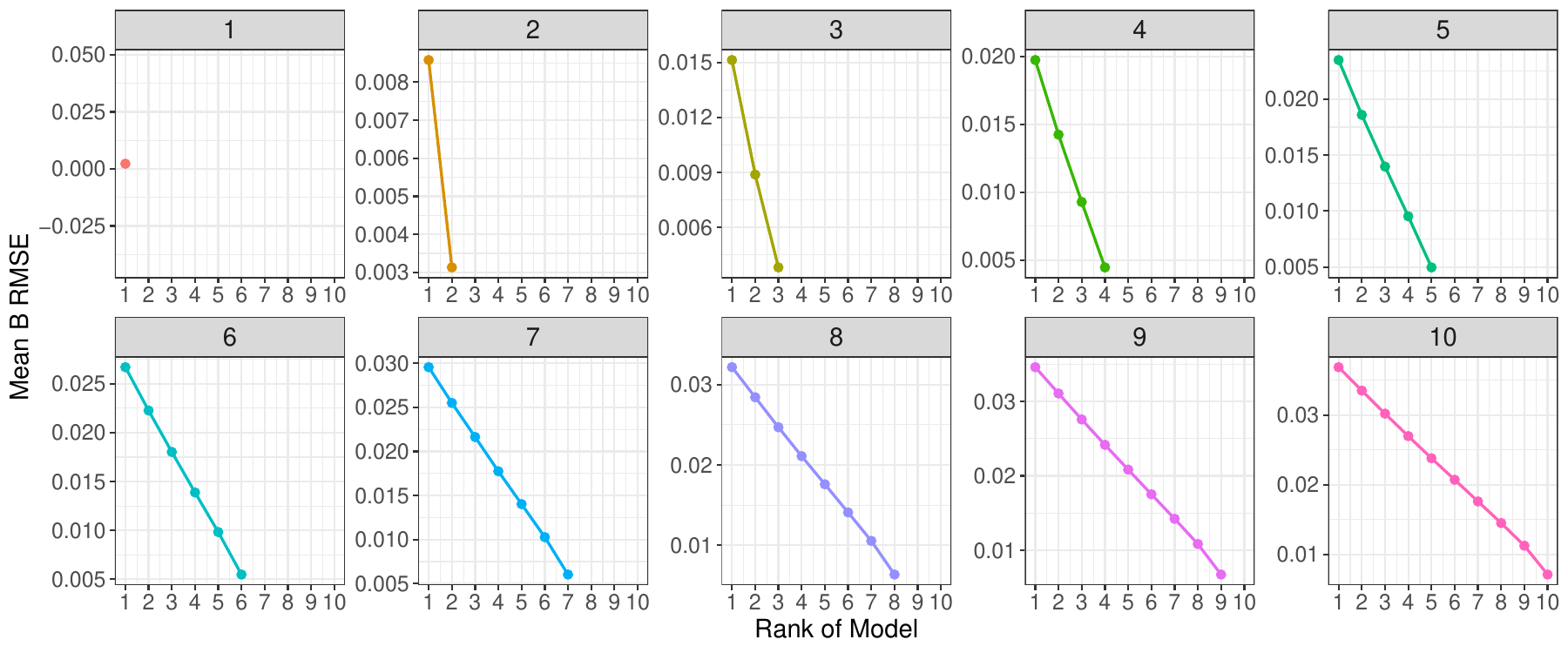}
    \caption{\textbf{RR-GBM estimation accuracy as a function of true rank and model rank.}
    Panels are labeled according to $N_0$, the true rank of $B$ used to simulate data. The x-axis is $N$, the rank of $B$ used to fit the data. 
    Each point is the average RMSE over 100 simulations, where for each simulation we compute the RMSE between the estimate $\hat{B} = \hat{Q}\hat{\Lambda}\hat{R}\T$ and the true matrix $B_0$. Negative Binomial data was generated with 1 latent factor ($M =1$). Each plot only goes up to $N = N_0$ since models with $N > N_0$ are degenerate and the estimation procedure does not converge.}
    \label{fig:true-reduced-rank-performance}
\end{figure}

\begin{figure}[H]
    \includegraphics[width=1\linewidth]{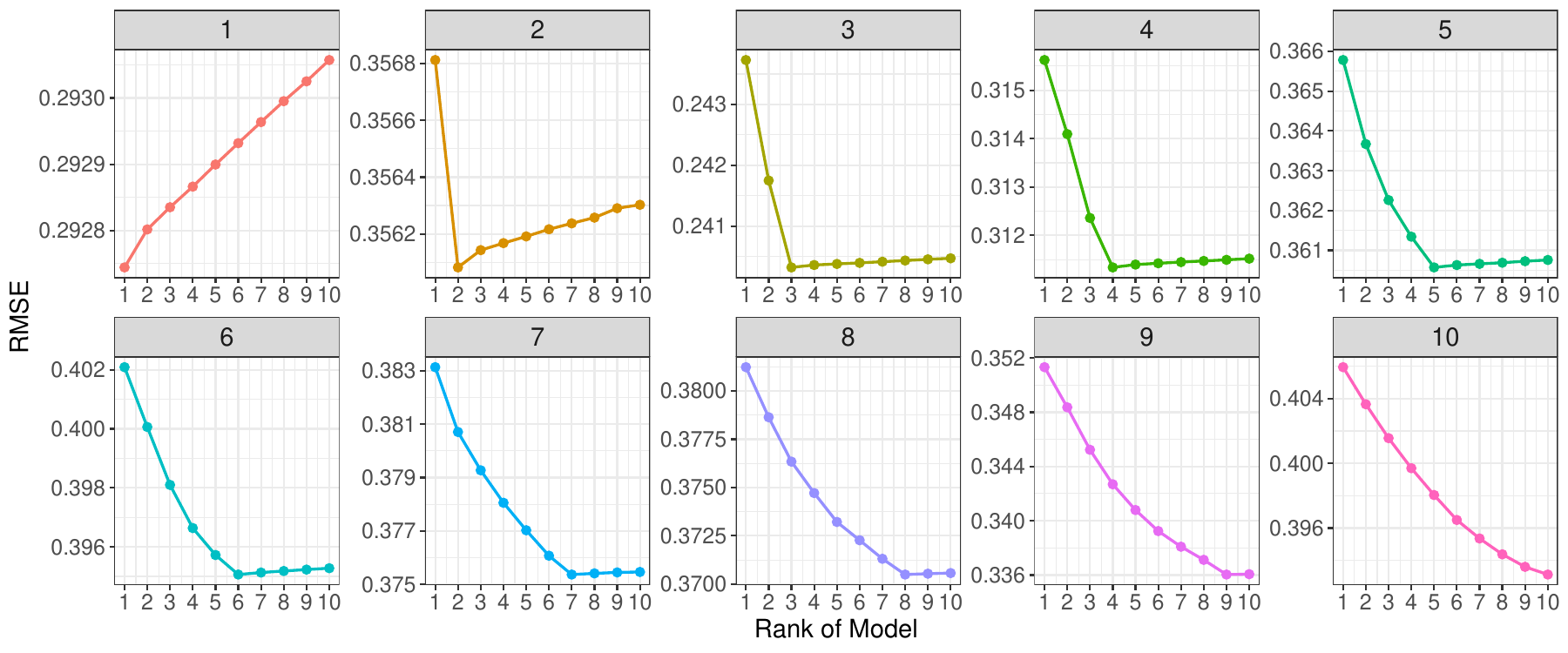}
    \caption{\textbf{Empirical selection of the rank in RR-GBM on a single replicate dataset.}
d    Panels are labeled according to the  true approximate rank of $B$, where the corresponding $L-N_{0}$ singular values are 0.01 times their randomly generated value to create a $B$ with approximate rank $N_{0}$. Each dot indicates the RMSE between the true values and predicted values  from one simulation run. Poisson data was generated and fit using a model with one latent factor ($M =1$).}
    \label{fig:empirical-rank-selection-single-run}
\end{figure}

\begin{figure}[H]
    \includegraphics[width=1\linewidth]{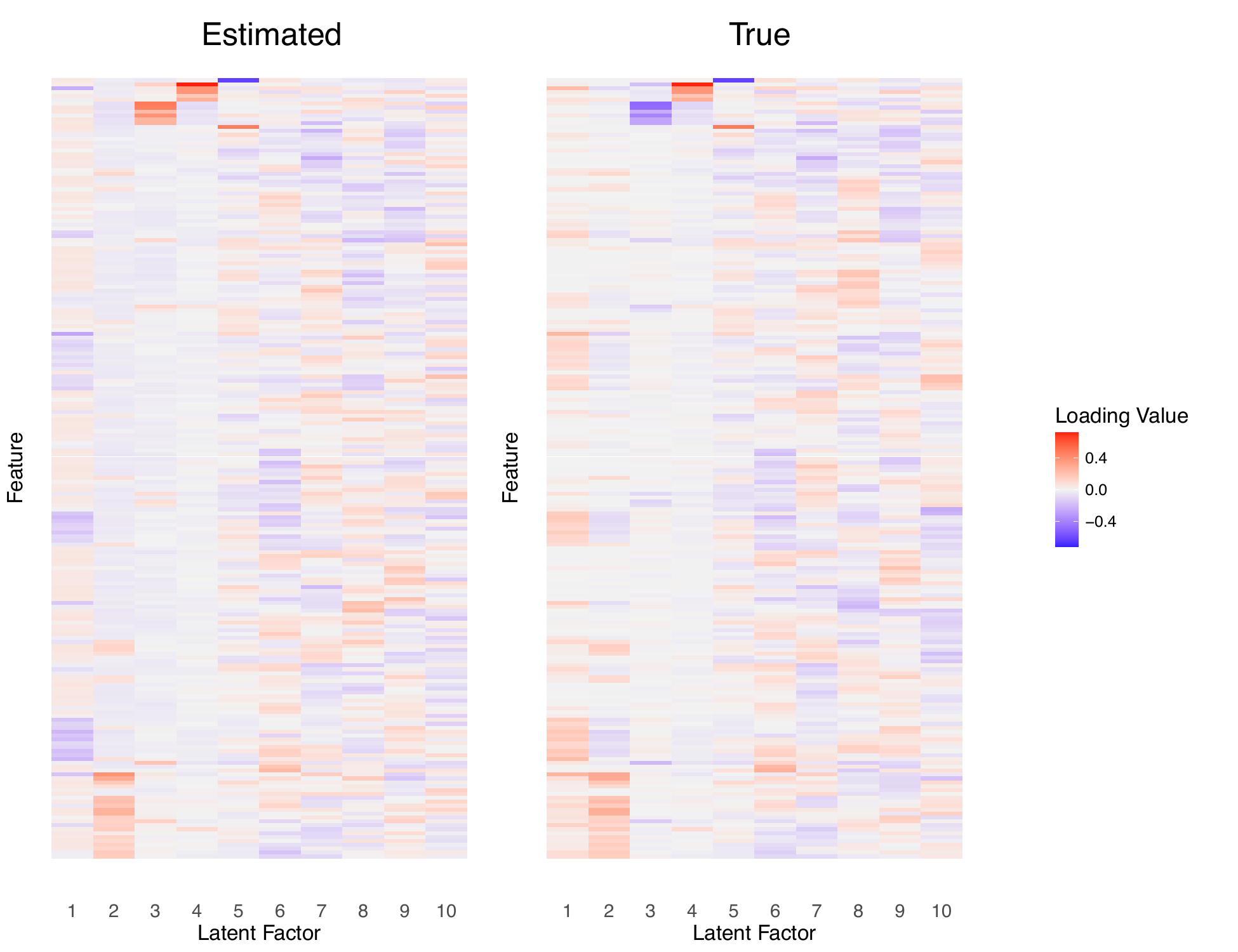}
    \caption{\textbf{Visualizing the effect of sample covariates on feature level latent factors in RR-GBM.}
    The estimate $\hat{Q}$ and true matrix $Q_{0}$ are shown. Data were simulated with 20,000 features, 200 samples, 1 latent factor, and 20 Bernoulli sample covariates with a Negative Binomial outcome distribution. The feature level coefficients for the five sample covariates 18, 11, 16, 3, 19, were altered at random with $\{1\%,\, 5\%,\, 10\%,\, 30\%,\, 50\%\}$ of feature coefficients being doubled, respectively. The true $B_{0}$ matrix was simulated as described in \cref{sec:sim-data-generation}, before the selected coefficients were altered. The rows are hierarchically clustered in the true $Q_{0}$ and the respective ordering is applied to the estimated $\hat{Q}$ matrix. All 20,000 features are plotted, with runs of features with similar loadings being blocked together by the limited resolution of the figure.}
    \label{fig:differential-effects-ground-truth-Q}
\end{figure}

\end{document}